# Modeling public mood and emotion: Twitter sentiment and socio-economic phenomena.

Johan Bollen
School of Informatics and Computing
Indiana University
Bloomington, IN
jbollen@indiana.edu

Alberto Pepe
Center for Embedded Networked Sensing
UCLA
Los Angeles, CA
apepe@ucla.edu

Huina Mao
School of Informatics and Computing
Indiana University
Bloomington, IN
huinmao@umail.iu.edu

## ABSTRACT

Microblogging is a form of online communication by which users broadcast brief text updates, also known as tweets, to the public or a selected circle of contacts. A variegated mosaic of microblogging uses has emerged since the launch of Twitter in 2006: daily chatter, conversation, information sharing, and news commentary, among others. Regardless of their content and intended use, tweets often convey pertinent information about their author's mood status. As such, tweets can be regarded as temporally-authentic microscopic instantiations of public mood state. In this article, we perform a sentiment analysis of all public tweets broadcasted by Twitter users between August 1 and December 20, 2008. For every day in the timeline, we extract six dimensions of mood (tension, depression, anger, vigor, fatigue, confusion) using an extended version of the Profile of Mood States (POMS), a well-established psychometric instrument. We compare our results to fluctuations recorded by stock market and crude oil price indices and major events in media and popular culture, such as the U.S. Presidential Election of November 4, 2008 and Thanksgiving Day. We find that events in the social, political, cultural and economic sphere do have a significant, immediate and highly specific effect on the various dimensions of public mood. We speculate that large scale analyses of mood can provide a solid platform to model collective emotive trends in terms of their predictive value with regards to existing social as well as economic indicators.

## Categories and Subject Descriptors

H.4.m [**Information Systems**]: Miscellaneous

## 1. INTRODUCTION

Microblogging is an increasingly popular form of communication on the web. It allows users to broadcast brief text updates to the public or to a limited group of contacts. Microblog posts, commonly known as *tweets*, are extremely short in comparison to regular blog posts: 140 characters or less. The launch of Twitter[1] in October 2006 is responsible for the popularization of this simple, yet vastly popular form of communication on the web. Following Twitter, microblogging capabilities have been implemented by many social networking websites in the form of status updates, e.g. on Facebook[2] and Myspace[3].

Users of these online communities use microblogging to broadcast different types of information. A recent analysis of the Twitter network revealed uses of microblogging for a) daily chatter, e.g., posting what one is currently doing, b) conversations, i.e., directing tweets to specific users in their community of followers, c) information sharing, e.g., posting links to web pages, and d) news reporting, e.g., commentary on news and current affairs [8].

Despite the diversity of uses emerging from such a simple communication channel, it has been noted that tweets normally tend to fall in one of two different content camps: users that microblog about themselves and those that use microblogging primarily to share information [15]. In both cases, tweets can convey information about the mood of their authors. In the former case, mood expressions are evident by an explicit "sharing of subjectivity" [4], e.g. "I am feeling sad"[4]. In other cases, even when someone is not specifically microblogging about their personal emotive status, the message can reflect their mood, e.g. "Colin Powell's endorsement of Obama: amazing. :)". As such, tweets may be regarded as microscopic instantiations of mood. It follows that the collection of all tweets published over a given time period can unveil changes in public mood state at a larger scale.

In this article, we explore how public mood patterns, as evidenced from a sentiment analysis of Twitter posts published between August 1 and December 20, 2008, relate to fluctuations in macroscopic social and economic indicators in the same time period. On the basis of a large corpus of public Twitter posts we look specifically at the interplay between a) macroscopic socio-economic events, such as a sharp drop in the stock market, a sudden rise in the oil price and the outcome of a political election, and b) the public's mood state measured by a well-established six-dimensional psychometric instrument. With our analysis we attempt to assess whether a quantifiable relationship between overall public mood and social, economic and other major events in the media and popular culture can be identified.

---

[1]Twitter — http://twitter.com
[2]facebook — http://www.facebook.com/
[3]myspace — http://www.myspace.com
[4]Please note that all the examples presented in this article are taken from the Twitter corpus analyzed. Whenever necessary they have been anonymized.

## 2. RELATED LITERATURE

An increasing number of empirical analyses of sentiment and mood are based on textual collections of public user-generated data on the web. Sentiment analysis and opinion mining are now a research domain of their own — sometimes referred to as "subjectivity analysis" — whose methods and applications were extensively surveyed in much detail by Pang and Lee [19].

Different methodological approaches have been used to extract sentiment from text. Some methods are grounded in natural language processing (NLP) and rely on word constructs (n-grams) found in text to extract sentiment towards a subject (favorable or unfavorable). NLP methods have been used to extract sentiment and opinion from texts such as camera [16] and pharmaceutical reviews [25]. Other techniques of sentiment analysis, rooted in machine learning, use supported vector machines (SVM) to classify text in positive or negative mood classes based on pre-classified training sets. SVM has been used to classify noisy customer feedback data [6] and movie reviews using a 5-point scale [18]. A number of hybrid methods that blend NLP and machine learning techniques have also appeared in the literature [22].

Besides the textual content discussed thus far (movie reviews, camera reviews, customer feedback), sentiment analyses have touched on many different kinds of personal online content. Personal websites such as blogs and online journals are often awash with emotive information and have been extensively used to deduct everyday happiness [12], explore trends and seasonality [1], forecast mood [13], and predict sales of books [7] and movies [14]. Some similar analytical tools operate entirely on the web: We Feel Fine[5] constantly harvests blog posts for occurrences of the phrases "I feel" and "I am feeling" and provides statistics and visualizations of past and current geo-tagged mood states. A similar online site, Moodviews[6], constantly tracks a stream of Livejournal weblogs that are user-annotated with a set of pre-defined moods.

The results generated via the analysis of such collective mood aggregators are compelling and indicate that accurate public mood indicators can be extracted from online materials. Using publicly available online data to perform sentiment analyses reduces enormously the costs, efforts and time needed to administer large-scale public surveys and questionnaires. These data and results present great opportunities for psychologists and social scientists. Yet, while blogs have largely been analyzed for mood patterns, not much research has yet addressed social networking sites and microblogging platforms. Recently, emotion has been extracted from public communication on Myspace [24] and status updates on Facebook[7], but we could not find any large scale sentiment analysis of Twitter, other than a study focused on microbloggers' response to the death of Michael Jackson [9]. This may be due to the fact that microblogging and social networking sites are fairly recent forms of online communication (at least when compared to blogs).

Scale may be an issue as well. Sentiment analysis techniques rooted in machine learning yield accurate classification results when sufficiently large data is available for testing and training. Minute texts such as microblogs may however pose particular challenges for this approach. In fact, the Twitter analysis of Jackson's death mentioned above was performed using a term-based matching technique based on the Affective Norms for English Words (ANEW). ANEW provides pre-existing, normed emotional ratings for nearly 3,000 terms along three dimensions (pleasure, arousal, dominance) [2] and has been recently employed to measure mood of song lyrics, blog posts and U.S. Presidents speeches [5]. Since it doesn't require training and testing, this syntactical approach may enable sentiment analysis for very small text data where machine learning techniques may not be appropriate.

In this study, we analyze the public's emotional state over a 5-month period by using a term-based emotional rating system known as the Profile of Mood States (POMS) and discussed in the next section. In previous work, we extended POMS making it suitable for sentiment analyses of online corpora. We applied and validated our extended version of POMS on a textual database of emails "sent to the future" (i.e., time capsules) inferring fluctuations in public mood states up to year 2036 [20].

## 3. METHODS

Our methodology consists of the following steps:

1. Definition of data and mood assessment instrument.
2. Data cleaning, parsing and normalization.
3. Time series production: aggregation of POMS mood scores over time.
4. Comparison of produced mood time series to socio-economic indicators.

In the following subsections we outline each of these steps in more details.

### 3.1 Data and Instruments

Our analysis is based on two data sources. First, we manually compile a timeline of events that took place during the time period analyzed in this article: August 1 to December 20, 2008. This timeline summarizes a number of macro- and micro-scopic social, economic, and political events, including:

- major fluctuations recorded by economic indices, such as gas prices and stock market indices;
- important international and U.S.-based political events;
- popular news stories, such as deaths of music and film icons;
- natural and man-made disasters, such as earthquakes and plane crashes.

This timeline, visualized at the very top of the master diagram of this article (Fig. 9), was constructed blending data from the Dow Jones Industrial Average (DJIA) and the West Texas Intermediate oil price (WTI) with a database of popular public events published on Wikipedia[8]. The following 2008 events were selected and are presented on the timeline:

---
[5]We Feel Fine — http://www.wefeelfine.org/
[6]Moodviews — http://moodviews.com
[7]United States Gross National Happiness — http://apps.facebook.com/usa_gnh/

---
[8]Wikipedia, 2008 — http://en.wikipedia.org/wiki/2008

**Aug 8 to 24** Summer Olympics take place in Beijing, China
**Aug 17** Swimmer Michael Phelps wins 8 gold medals at Summer Olympics
**Aug 20** Spanair Flight 5022 crashes at Madrid Barajas Airport killing 154
**Aug 25 to 28** Democratic National Convention (DNC)
**Aug 29** John McCain chooses Sarah Palin as running mate
**Sep 1 to 4** Republican National Convention (RNC)
**Sep 15** Lehman Brothers files for Chapter 11 bankruptcy protection
**Sep 22** WTI oil price skyrockets to $120
**Sep 26** First presidential debate before U.S. elections; Paul Newman, American actor, dies
**Oct 6** DJIA drops below 10,000 points
**Oct 9** Major banking and financial crisis in Iceland
**Nov 4** United States presidential election; Barack Obama is elected the 44th President of the United States
**Nov 21** DJIA drops nearly 700 points; the U.S. is officially in economic recession
**Nov 25** Major political crisis in Thailand: protesters occupy Bangkok Airport
**Nov 27** Thanksgiving day
**Dec 1** Consumer Confidence Index (CCI) drops to 38.6, the lowest value ever recorded
**Dec 11** Bernard Madoff arrested in enormous fraud
**Dec 16** Fed cuts rates to a record low rate

Second, we collected all public tweets broadcasted by Twitter users in the same time period, i.e. August 1 to December 20, 2008. The resulting data set consists of 9,664,952 tweets temporally distributed as shown in Fig.1.

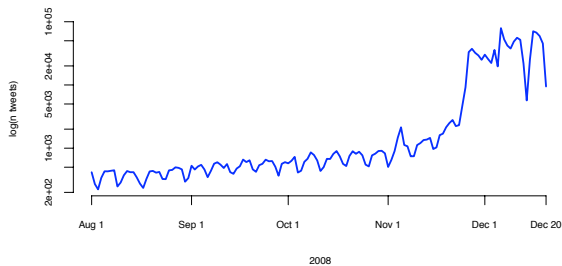

**Figure 1:** Daily tweet distribution: August 1 to December 20, 2008.

We perform a sentiment analysis of the Twitter data using an extended version of a well established psychometric instrument, the Profile of Mood States (POMS) [11]. POMS measures six individual dimensions of mood, namely *Tension*, *Depression*, *Anger*, *Vigour*, *Fatigue*, and *Confusion*. The POMS is not intended for large-scale textual analysis. Rather, it is a psychometric questionnaire composed of 65 base terms. Respondents to the questionnaire are asked to indicate on a five-point intensity scale how well each one of the 65 POMS adjectives describes their present mood state. The respondent's ratings for each mood adjective are then transformed by means of a scoring key to a 6-dimensional mood vector. The POMS is an easy-to-use, low-cost instrument whose factor-analytical structure has been repeatedly validated, recreated, and applied in hundreds of studies [17, 10]. A number of reduced versions have also appeared in specialized literature, with the aim to condense the number of test terms and thus to reduce the time and effort on the part of human subjects to complete the POMS questionnaire [23, 3].

In previous work [20], we have introduced and validated an extended version of the POMS, not intended to be administered as a questionnaire to human subjects, but rather to be applicable to large textual corpora. The instrument we created, POMS-ex, extends the original set of 65 mood adjectives to 793 terms, including synonyms and related word constructs extracted from the Princeton University's WordNet (version 3.0) and Roget's New Millennium Thesaurus (First Edition).

### 3.2 Text cleaning and parsing

Each individual tweet in our Twitter collection of 9,664,952 tweets is normalized and parsed before processing as follows:

1. Separation of individual terms on white-space boundaries
2. Removal of all non-alphanumeric characters from terms, e.g. commas, dashes, etc.
3. Conversion to lower-case of all remaining characters.
4. Removal of 214 standard stop words, including highly common verb-forms.
5. Porter stemming [21] of all remaining terms in tweet.

Thus, each tweet is converted to an ordered set of terms, filtered for stopwords and non-alphanumeric characters, converted to lower-case, and Porter-stemmed. For example, the tweet "Feeling too lazy to go to the shops and get something to eat" is normalized to the term set {feel, lazi, shop, someth, eat}. In addition, since we are interested only in those tweets that represent an explicit expression of individual sentiment, or those that reflect an individual's present status, we only retain tweets that match the following regular expressions "feel", "I'm", "Im", "am", "am", "being", and "be". In order to avoid spam messages and other information-oriented tweets, we also filter out tweets that match the regular expressions "http:" or "www.". The diversity of expression on Twitter is tremendous and few tweets directly express or pertain to users' mood or sentiment. Tweets conforming to these particular patterns are a minority and consequently this procedure leads to a reduction of the 9,664,952 tweets to a set of 1.1M tweets containing mostly expressions of individual mood states.

### 3.3 POMS scoring

Each tweet in the resulting sub-set of 1.1M normalized tweets is then POMS-scored, i.e. the POMS-scoring function $\mathcal{P}(t)$ maps each tweet to a six-dimensional mood vector $m \in \mathbb{R}^6$. The entries of $m$ represent the following six dimensions of mood, namely *Tension*, *Depression*, *Anger*, *Vigour*, *Fatigue*, and *Confusion*.

The POMS-scoring function $\mathcal{P}(t)$ simply matches the terms extracted from each tweet to the set of POMS mood adjectives for each of POMS' 6 mood dimensions. Each tweet $t$ is represented as the set $w$ of $n$ terms. The particular set of $k$ POMS mood adjectives for dimension $i$ is denoted as the set $p_i$. The POMS-scoring function, denoted $\mathcal{P}$, can thus be defined as follows:

$\mathcal{P}(t) \rightarrow m \in \mathbb{R}^6 =$
$[||w \cap p_1||, ||w \cap p_2||, ||w \cap p_3||, ||w \cap p_4||, ||w \cap p_5||, ||w \cap p_6||]$

The resulting mood vector $m$ for tweet $t$ is then normalized to produce the unit mood vector

$$\hat{m} = \frac{m}{||m||}$$

.

## 3.4 Time series production and normalization

We produce an aggregate mood vector $m_d$ for the set of tweets submitted on a particular date $d$, denoted $T_d \subset T$ by simply averaging the mood vectors of the tweets submitted that day, i.e.

$$m_d = \frac{\sum_{\forall t \in T_d} \hat{m}}{||T_d||}$$

The time series of aggregated, daily mood vectors $m_d$ for a particular period of time $[i, i+k]$, denoted $\theta_{m_d}[i, k]$, is then defined as:

$$\theta_{m_d}[i,k] = [m_i, m_{i+1}, m_{i+2}, \cdots, m_{i+k}]$$

A different number of tweets is submitted on any given day. Each entry of $\theta_{m_d}[i, k]$ is therefore derived from a different sample of $N_d = ||T_d||$ tweets. The probability that the terms extracted from the tweets submitted on any given day match the given number of POMS adjectives $N_p$ thus varies considerably along the binomial probability mass function:

$$P(K = n) = \binom{N_p}{||W(T_d)||} p^{||W(T_d)||}(1-p)^{N_p - ||W(T_d)||}$$

where $P(K = n)$ represents the probability of achieving $n$ number of POMS term matches, $||W(T_d)||$ represents the total number of terms extracted from the tweets submitted on day $d$ vs. $N_p$ the total number of POMS mood adjectives. Since the number of tweets per day has increased consistently from Twitter's inception in 2006 to present, this leads to systemic changes in the variance of $\theta_{m_d}[i, k]$ over time, as shown in Fig. 2. In particular, the variance is larger in the early days of Twitter, when tweets are relatively scarce. As the number of tweets per day increases, the variance of the time series decreases. This effect makes it problematic to compare changes in the mood vectors of $\theta[i, k]$ over time.

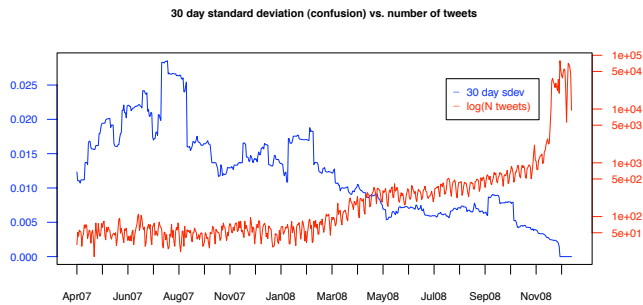

**Figure 2:** Standard deviation values of POMS Confusion scores within a 30 day window vs. the number of tweets submitted.

For this reason, we convert all mood values for a given day $i$ to z-scores so that they would be normalized with respect to a local mean and standard deviation observed within the period $[i-k, i+k]$, i.e. a sliding window of $k$ days before and after the particular date. The z-score of a mood vector $\hat{m}_i$ for date $i$, denoted $\tilde{m}_i$, is then defined as:

$$\tilde{m}_i = \frac{\hat{m}_i - \bar{x}(\theta[i, \pm k])}{\sigma(\theta[i, \pm k])}$$

where $\bar{x}(\theta[i, \pm k])$ and $\sigma(\theta[i, \pm k])$ represent the mean and standard deviation of the time series within the local $[i, \pm k]$ days neighborhood of $\hat{m}_i$. When combined, the normalized mood vectors form the normalized time series:

$$\tilde{\theta}_{m_d}[i,k] = [\tilde{m}_i, \tilde{m}_{i+1}, \tilde{m}_{i+2}, \cdots, \tilde{m}_{i+k}]$$

The effect of the z-score normalization is shown in Fig. 3 for the time series of the POMS confusion dimension over the course of 600 days[9]. Where the time series is produced from a small number of tweets resulting in large swings in the un-normalized mood vectors, the magnitude of these swings is reduced by a commensurately high standard deviation. Where the time series is based on a larger sample of tweets, the standard deviation is smaller and thus a smaller swing in the un-normalized mood vectors is required to produce significant z-score fluctuations. As a result, the normalized time series fluctuates around a mean of zero and its fluctuations are expressed on a common scale, namely the standard deviation regardless of the number of tweets submitted on a particular date. This allows us to interpret the magnitude of the time series' fluctuations in terms of a common scale.

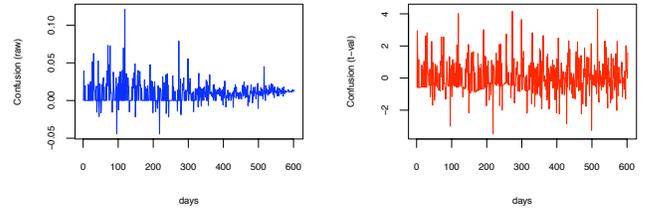

**Figure 3:** Raw POMS Confusion scores (left) vs. their z-scores normalization (right).

Thus, the normalized time series highlights local, short-term deviations; its values fluctuate around mean of zero. However, to asses long-term shifts in mood levels, we adopt a variable mean so that we can make comparisons of general mood levels between different periods of time while maintaining normalized variance; we define the variance-normalized mood vector $\bar{m}_i$ as follows:

$$\bar{m}_i = \frac{\hat{m}_i}{\sigma(\theta[i, \pm k])}$$

and consequently we obtain the following variance-normalized time series:

$$\bar{\theta}_{m_d}[i,k] = [\bar{m}_i, \bar{m}_{i+1}, \bar{m}_{i+2}, \cdots, \bar{m}_{i+k}]$$

The effects of variance-normalization are shown in Fig. 4.

The above described normalizations result in two different timeseries for the same period, from August 1 to December 20, 2008:

1. Z-score normalized: a 153 day, 6-dimensional time series that fluctuates around a mean of zero on a scale of

---

[9]The graph is generated over a wider time frame than the period August 1, 2008 to December 20, 2008 under investigation to better illustrate this effect

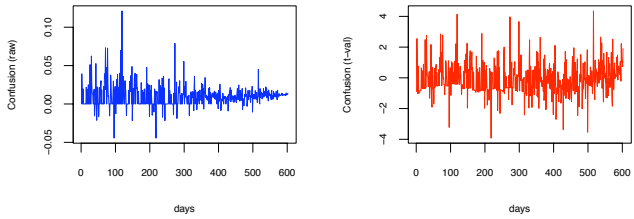

**Figure 4: Raw POMS Confusion scores (left) vs. their variance normalization (right).**

1 standard deviation. This is used to highlight short-term fluctuations of public mood as a result of particular short-term events;

2. Variance normalized: a 153 day, 6-dimensional time series whose variance has been normalized to a scale of 1 standard deviation. This is used to assess changing mood levels over time in relation to long-term changes in socio-economic indicators.

The results of our data collection, aggregation and time series production outlined above are summarized in the master diagram of Fig. 9. Starting from the top, Fig. 9 displays for the period under study:

1. a timeline of the most important social, cultural, political and economic events;
2. the DJIA and WTI trend lines;
3. the time series extracted from our collection of tweets for each of the POMS mood dimensions, z-score normalized.

Shaded areas indicate the span of events that lasted for more than one day. Vertical lines originate in the time line's events and run across all mood dimensions to provide a visual frame of reference.

## 4. RESULTS

Our investigation of the produced public mood time series proceeds in two phases. First, we assess the validity of our sentiment analysis by examining the effects of particular events, namely the U.S. Presidential election of November 4, 2008, and the Thanksgiving holiday in the U.S., on our time series. Second, we examine the long-term effects of socio-economic indicators on general mood levels across longer periods of time.

### 4.1 Case studies

Our first case study is the 2008 US Presidential election which was held on November 4, 2008. The mood curves in Fig. 5 are presented as blue "sparklines" for each of the mood dimensions. The x-axis expresses time in days; it spans 15 days before and after election day. The period two days before and after election day is shown as a gray area for convenient location of mood changes in that period of time. The y-axis corresponds to mood z-scores, expressed in standard deviations from the mean. A scale is not provided since we are mostly interested in the pattern of increasing and decreasing POMS mood scores for each of the different dimension, rather than their exact value. However, all discussed peaks and troughs are nearly or above 2 standard deviations from the mean as shown in Fig. 9.

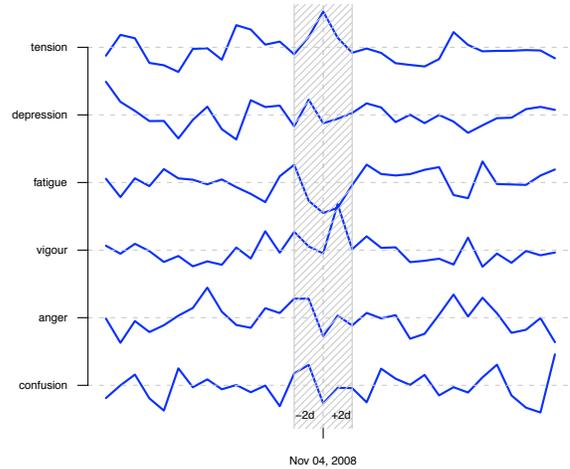

**Figure 5: Sparklines for public mood before, during and after the US presidential election on November 4, 2008.**

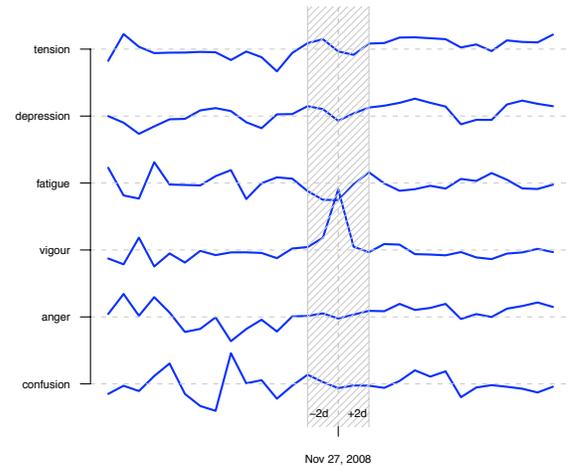

**Figure 6: Sparklines for public mood before, during and after Thanksgiving on November 27, 2008.**

The mood curves shown in Fig. 5 provide a fine-grained view of public mood changes in the three-day period surrounding election day (November 4, 2008). We observe a spike in Depression and Confusion on November 3, and remarkably a sharp drop in Fatigue that started two days before election day. This could indicate a surge in tweets that express doubt and apprehension about the outcome of the election, and calls for action on election day which leads to a drop in Fatigue. November 4 is characterized by a drop in Confusion to baseline levels, a further drop in Fatigue and a sharp peak in Tension, indicating tweets that express calls for action and concern and/or excitement over the election. The outcome of the election is celebrated on November 5 where mood levels drop to nominal levels, except a significant spike in Vigour and a large drop in Fatigue. An examination of tweet content reveals a preponderance of tweets expressing high levels of energy and positive sentiments over the outcome of the election[10].

Our second case study relates to the celebration of Thanksgiving (November 27), a national holiday in the U.S. that is nearly always associated with copious amounts of calorie-dense food, family gatherings and American football. The sparklines shown in Fig. 6 bear this out. All mood dimensions remain nearly at baseline levels with the exception of Vigour which spikes significantly on Thanksgiving Day indicating happy, active mood. We also notice a dip in Fatigue which along with the significant increase in Vigour further confirms the picture of Thanksgiving as a happy, energetic holiday.

The sparklines in Fig. 5 and Fig. 6 do not do justice to the large magnitudes of the discussed mood changes, however. Against the backdrop of the week- or month-long patterns as shown in Fig. 9 the spikes in Vigour and Tension surrounding the presidential election reflect a move of nearly 4 standard deviations, respectively -1 to +3 standard deviations for Vigour and -2 to +2 standard deviations for Tension. Thanksgiving corresponds to the most significant positive spike in Vigour of the entire period we study, i.e. 0 to +4 standard deviations.

### 4.2 General correlation drivers versus public mood trends

In this sub-section, we examine the ability of large-scale economic indicators such as the DJIA and the WTI to drive public mood. In Fig. 9 we visualize the time series of the produced POMS dimensions of mood as well as the DJIA and WTI over the same period of time, namely August 1, 2008 to December 20, 2008.

Public sentiment fluctuated significantly in this tumultuous period under the influence of the U.S. presidential campaign and election, the failures of several large, international banks, the DJIA dropping in value from above 11,000 points to less than 9,000, significant changes in the price of crude oil, and the official start of the deepest world-wide economic recession since World War II. This is reflected by the large fluctuations of the mood curves shown in Fig. 9 which exhibit large swings in value that range from several standard deviations below the mean to several standard deviations above the mean on a daily or weekly scale. A few notable examples:

**August 17-20** Depression swings from -1 standard deviation to +3.3 standard deviations, and back in less than 3 days.

**August 28-September 2** Right after John McCain announces Sarah Palin as his running mate, Tension swings from -2 standard deviations to +2 standard deviations in a few days.

**October 20** Depression swings from -1 standard deviation to +2 standard deviations and back to -1.5 in the span of 3 or 4 days.

Throughout this tumult, the emotional response of the Twitter community was highly differentiated. None of the mood dimensions' values were statistically significantly correlated across all days in the period under investigation.

---

[10] Although the election results become known later in the evening of November 4, the date and time in our data are recorded in GMT+0. As a result even the immediate reactions to Obama's victory were mostly recorded on November 5 in our data.

We calculate pairwise Spearman Rank order correlations between each mood dimension by the day, thereby producing the 6×6 correlation matrix $M$ which contains no statistically significant correlations for $N = 141$.

$$M = \begin{bmatrix} \text{Ts} & \text{Cf} & \text{Vg} & \text{Ft} & \text{Ag} & \text{Dp} \\ 1.00 & 0.00 & 0.02 & -0.05 & 0.09 & 0.07 \\ 0.00 & 1.00 & -0.04 & 0.00 & 0.06 & -0.02 \\ 0.02 & -0.04 & 1.00 & -0.02 & 0.00 & -0.01 \\ -0.05 & 0.00 & -0.02 & 1.00 & -0.06 & -0.01 \\ 0.09 & 0.06 & 0.04 & -0.06 & 1.00 & 0.00 \\ 0.07 & -0.02 & -0.01 & -0.01 & 0.00 & 1.00 \end{bmatrix}$$

To assess the effect of changes in the DJIA and WTI on public mood levels, we define 4 crucial periods in which the DJIA underwent significant changes in value. We examine the extent of mood changes across those 4 periods. The following four periods were selected on the basis of the data shown in Fig. 9:

**DJIA-I: August 1 to 24** The Dow Jones remained stable at a value above 11,000.

**DJIA-II: September 15 to October 9** The DJIA falls precipitously from a value above 11,000 to less than 9,000.

**DJIA-III: October 9 to 25** A plateau is reached after the crash and the collapse of the Iceland banking system.

**DJIA-IV: December 1-20** : After Thanksgiving, the DJIA maintains a low plateau at 8500 to 9,000 points.

Fig. 7 shows the sparklines for the six mood dimensions as observed in the period under study. The displayed values are variance-normalized as discussed in Section 3.4, i.e. they are normalized according to a 30-day running standard deviation, but not their mean. This ensures the visibility of long-term trends in average mood levels over time. The DJIA periods as defined above are marked as gray bars on the graph.

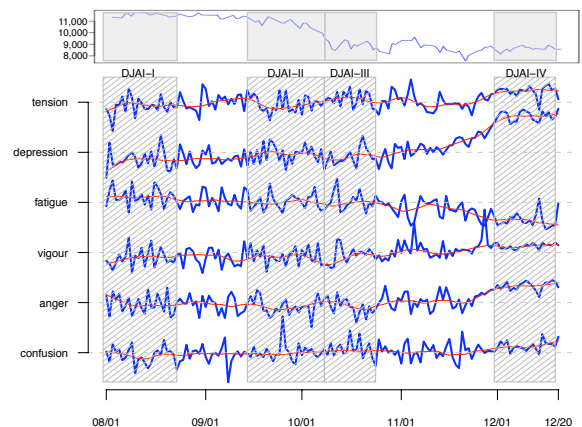

Figure 7: Sparklines for public mood in period August, 2008 to December 20, 2008 compared to DJIA values in 4 distinct periods of change.

The sparklines of Fig. 7 indicate significant changes in mood states over time. In particular, we observe a significant increase in Depression and a significant decrease of Fatigue between DJIA-III and DJIA-IV. In addition we see a significant uptick in Anger in that same period. Throughout the entire period under study we see increasing levels of Vigour and a slight increase in Tension. The mentioned mood curves either stabilize within DJIA-IV or increase at a lower rate. DJIA-I, DJIA-II and DJIA-III do not seem to be associated with significant increases in any of the POMS mood levels, however structural changes do seem to occur. Tension, Depression and Vigour seem to fluctuate more significantly within DJIA-II and DJIA-III, cf. the rapid oscillations of Depression early in DJIA-II.

Although visually indicative of significant public mood changes related to changes in the DJIA, we need to determine whether the perceived differences in mood levels are in fact statistically significant. We perform a Mann-Witney U-test over all 36 possible combinations of the time series observed within the 4 DJIA periods for each mood dimension, i.e. 4 DJIA periods $\sim$ 6 unique 2-way permutations x 6 mood dimensions = 36 Mann-Witney U-test. Of these comparisons, we report in Table 1 only those that resulted in $p < 0.05$.

| Dimension | DJIA periods | diff. median | p-value |
|---|---|---|---|
| Tension | I -III | (-0.473, 0.408) | 0.045 |
| Tension | I-IV | (-0.473, 1.466) | $< 0.0001$ |
| Tension | II-IV | (-0.435, 1.466) | $< 0.0001$ |
| Tension | III-IV | (0.408, 1.466) | 0.001 |
| Confusion | I -III | (-0.253, 0.28) | 0.032 |
| Confusion | I-IV | (-0.253, 0.751) | $< 0.0001$ |
| Confusion | II-IV | (-0.177, 0.751) | $< 0.0001$ |
| Vigour | I-IV | (-0.662, 0.776) | $< 0.0001$ |
| Vigour | II-IV | (-0.502, 0.776) | $< 0.0001$ |
| Vigour | III-IV | (-0.33, 0.776) | $< 0.0001$ |
| Fatigue | I-II | (0.561, -0.087) | 0.034 |
| Fatigue | I-IV | (0.561, -2.38) | $< 0.0001$ |
| Fatigue | II-IV | (-0.087, -2.38) | $< 0.0001$ |
| Fatigue | III-IV | (0.05 , -2.38) | $< 0.0001$ |
| Anger | I-IV | (-0.198, 2.043) | $< 0.0001$ |
| Anger | II-IV | (-0.458, 2.043) | $< 0.0001$ |
| Anger | III-IV | (-0.498, 2.043) | $< 0.0001$ |
| Depression | I-IV | (-0.983, 4.532) | $< 0.0001$ |
| Depression | II-IV | (-0.052, 4.532) | $< 0.0001$ |
| Depression | III-IV | (-0.825, 4.532) | $< 0.0001$ |

Table 1: **Statistically significant differences ($p < 0.05$) in mood levels between DJIA-I, DJIA-II, DJIA-III, DJIA-IV in each mood dimension.**

As Table 1 indicates, all mood curves undergo statistically significant changes from one DJIA period to the next. Most trends observed in Fig. 7 are thus highly statistically significant. For example, Depression mood levels change from a median of -0.983 standard deviations in DJIA-I to 4.532 in DJIA-IV ($p < 0.0001$). An even larger difference is observed for DJIA-II vs. DJIA-IV, i.e. -0.052 to 4.532 standard deviations ($p < 0.0001$. Similarly, Anger mood levels swing from a median of -0.198 to 2.043 ($p < 0.0001$) across DJIA-I and DJIA-IV, most of the change situated in the period before and during DJIA-IV. Tension experiences a similar shift. Fatigue on the other hand undergoes a continuous and significant decline across all DJIA periods.

We see significantly higher levels of Tension, Depression and Anger in DJIA-IV than in any of previous periods. This increase in negative sentiment occurs well after the DJIA has completed its fall to a stable, but relatively low plateau in comparison to its summer value. Depression however increases throughout the entire 3 month period. The observed movement of negative sentiment from period 1 to period 4 is thus opposite of the movement of the DJIA, but with most of the changes occurring in DJIA -IV.

A similar approach was adopted to determine whether WTI prices had an effect on public mood. Based on WTI values as shown in Fig. 9 we compared public mood in the following periods:

**WTI-I: August 1 to 22** WTI is above $100.
**WTI-II: September 15 to September 31** A brief spike followed by a decline.
**WTI-III: October 1 to November 21** Consistent drop from about $90 to $60.
**WTI-IV: November 22 to December 16** Slowing decline and relative stabilization.

In Fig. 8 the described WTI periods are superimposed on the same mood curves as shown in Fig. 7.

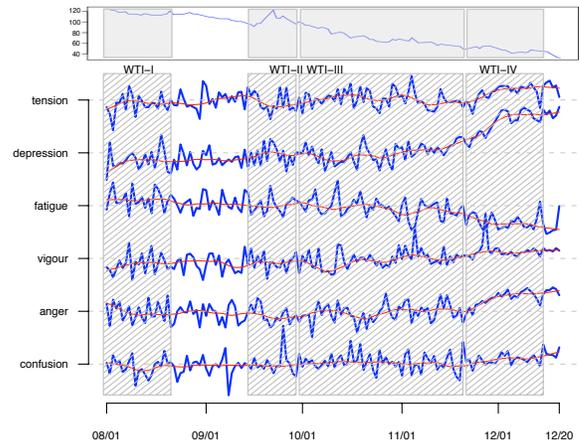

Figure 8: **Sparklines for public mood in period August 1, 2008 to December 20th, 2008 compared to WTI values in 4 distinct periods of change.**

The WTI periods overlap to some degree with the DJIA periods since they follow similar downward trajectories and could thus be expected to correspond to similar changes in mood level. WTI values however deviate from the downwards trajectory of the DJIA at one point in time, i.e the significant spike in WTI values from September 15 to September 31. At the start of WTI-II we do in fact see a set of rapid fluctuations of Depression and Vigour which however do not seem to affect median mood levels for this period. The boundary between WTI-III and WTI-IV is marked, much like the boundary between DJIA-III and DJIA-IV, with a rapid acceleration in the increase of Depression and Anger levels. In fact, the boundary between WTI-III and WTI-IV seems to correspond better to discontinuities in these mood curves than the boundary between DJIA-III and DJIA-IV. Table 2 list the p-values of statistically significant changes in mood levels across the various WTI periods. It contains more statistically significant differences for Vigour and Fatigue than those listed in Table 1. This could be construed as a possible indication that changes in WTI values better correspond to changes in mood levels in a temporal sense but act on different mood dimensions than DJIA values.

| Dimension | WTI periods | diff. median | p-value |
| --- | --- | --- | --- |
| Tension | I - III | (-0.473, -0.092) | 0.029 |
| Tension | I - IV | (-0.473, 1.113) | < 0.0001 |
| Tension | II - IV | (-0.773, 1.113) | 0.001 |
| Tension | III - IV | (-0.092, 1.113) | < 0.0001 |
| Confusion | I - IV | (-0.127, 0.541) | 0.001 |
| Confusion | II - IV | (-0.213, 0.541) | 0.005 |
| Vigour | I - III | (-0.813, -0.09) | 0.017 |
| Vigour | I - IV | (-0.813, 0.727) | < 0.0001 |
| Vigour | II - IV | (-0.463, 0.727) | < 0.0001 |
| Vigour | III - IV | (-0.09, 0.727) | < 0.0001 |
| Fatigue | I - II | (0.691, 0.122) | 0.039 |
| Fatigue | I - III | (0.691, -0.11) | 0.001 |
| Fatigue | I - IV | (0.691, -1.905) | < 0.0001 |
| Fatigue | II- IV | (0.122, -1.905) | < 0.0001 |
| Fatigue | III - IV | (-0.11, -1.905) | < 0.0001 |
| Anger | I - IV | (-0.482, 1.55) | < 0.0001 |
| Anger | II- IV | (-0.79, 1.55) | < 0.0001 |
| Anger | III - IV | (-0.279, 1.55) | < 0.0001 |
| Depression | I - III | (-0.983, 0.032) | 0.003 |
| Depression | I - IV | (-0.983, 3.696) | < 0.0001 |
| Depression | II- IV | (0.017, 3.696) | < 0.0001 |
| Depression | III - IV | (0.032, 3.696) | < 0.0001 |

Table 2: Statistically significant differences ($p < 0.05$) in mood levels between WTI-I, WTI-II, WTI-III, WTI-IV in each mood dimension.

Similar to the DJIA period, increases in Anger, Tension and Depression occur well after the decreases in WTI values have moderated. Long-term fluctuations in macro-economic indicators such as the DJIA and WTI may thus have a more delayed, cumulative effect, whereas the high variability of public mood in our collection of tweets seems to be predominantly driven by short-term events such as the news cycle, elections, national holidays, and other events. The increased levels of Anger, Tension and Depression in the winter of 2008 may in fact be the result of an accumulation of bad economic news throughout the Fall as the economic recession deepened and a significant fraction of the Twitter community was increasingly feeling disenfranchised. The rise in Anger and Depression in late 2008 may also signal the vanguard of what appears to be a populist movement in opposition to the new Obama administration.

In summary, although inconclusive with regards to the relation between long-term changes in economic indicators, our results seem to indicate at least the following. First, events in the social, political, cultural and economical sphere do have a significant, immediate and highly specific effect on the various dimensions of public mood. These effects are short-lived as could be expected for mood states that are ephemeral and variable by definition. Second, economic events do seem to have an effect on public mood, but only to the degree that they correspond to rapid changes of economic indicators such as the DJIA. Long-term changes seem to have a more gradual and cumulative effect. Third, continued negative drivers seem to have an effect on public mood but this effect may be manifested by short bursts of negative sentiment such as those observed on October 20, 2008. Finally, we would like to speculate that the social network of Twitter may highly affect the dynamics of public sentiment. Although we do not investigate the Twitter subscription network in this article, our results are suggestive of escalating bursts of mood activity, suggesting that sentiment spreads across network ties.

## 5. CONCLUSION

In this article, we perform a sentiment analysis of tweets (brief text updates) publicly broadcasted on the Twitter microblogging platform between August 1 and December 20, 2008. What makes Twitter an interesting data source to explore public mood trends is the very flexible and ephemeral nature of its content. Twitter is currently used for daily chatter, open conversation, information sharing, public opinion, and political and news commentary. A number of these uses are a concrete example of user appropriation on the web: Twitter was initially intended to enable users to simply publish short messages answering the question "what are you doing?" but it has quickly grown beyond this initial use case. The so-called Green Revolution following the 2009 Iranian presidential elections is a representative example of this: a political protest organized, aided, and promoted by social exchange on microblogging platforms. Another crucial characteristic of Twitter content is its timeliness. Tweets are so brief that they are necessarily associated with a specific moment (the tweet's timestamp). As such, tweets relate to a much narrower temporal window than less ephemeral user-generated texts, such as blogs and journals.

In this article, we speculate that these two characteristics — the diversity of use scenarios and the ephemeral nature of microblogging — make Twitter an interesting source to investigate public mood and emotive trends. We regard an individual tweet as a microscopic, temporally-authentic instantiation of sentiment. We measure this sentiment using a syntactic, term-based approach, in order to detect as much mood signal as possible from very brief Twitter messages. We use the extended version of a well-established psychometric instrument, the Profile of Mood States (POMS). For each tweet published between August 1 and December 20, 2008, we extract a six-dimensional vector representing the tweet's mood. We aggregate mood components on a daily scale and compare our results to the timeline of cultural, social, economic, and political events that took place in that time period. In particular, we compare mood trends to fluctuations recorded by stock market and crude oil price indices and to major events in media and popular culture, such as the U.S. Presidential Election of November 4, 2008 and Thanksgiving Day. We find that social, political, cultural and economic events are correlated with significant, even if delayed fluctuations of public mood levels.

With the present investigation, we bring about the following methodological contribution: we argue that sentiment analysis of minute text corpora (such as tweets) is efficiently obained via a syntactic, term-based approach that requires no training or machine learning. Moreover, we stress the importance of measuring mood and emotion using well-established instruments rooted in decades of empirical psychometric research. Finally, we speculate that collective emotive trends can be modeled and predicted using large-scale analyses of user-generated content but results should be discussed in terms of the social, economic, and cultural spheres in which the users are embedded.

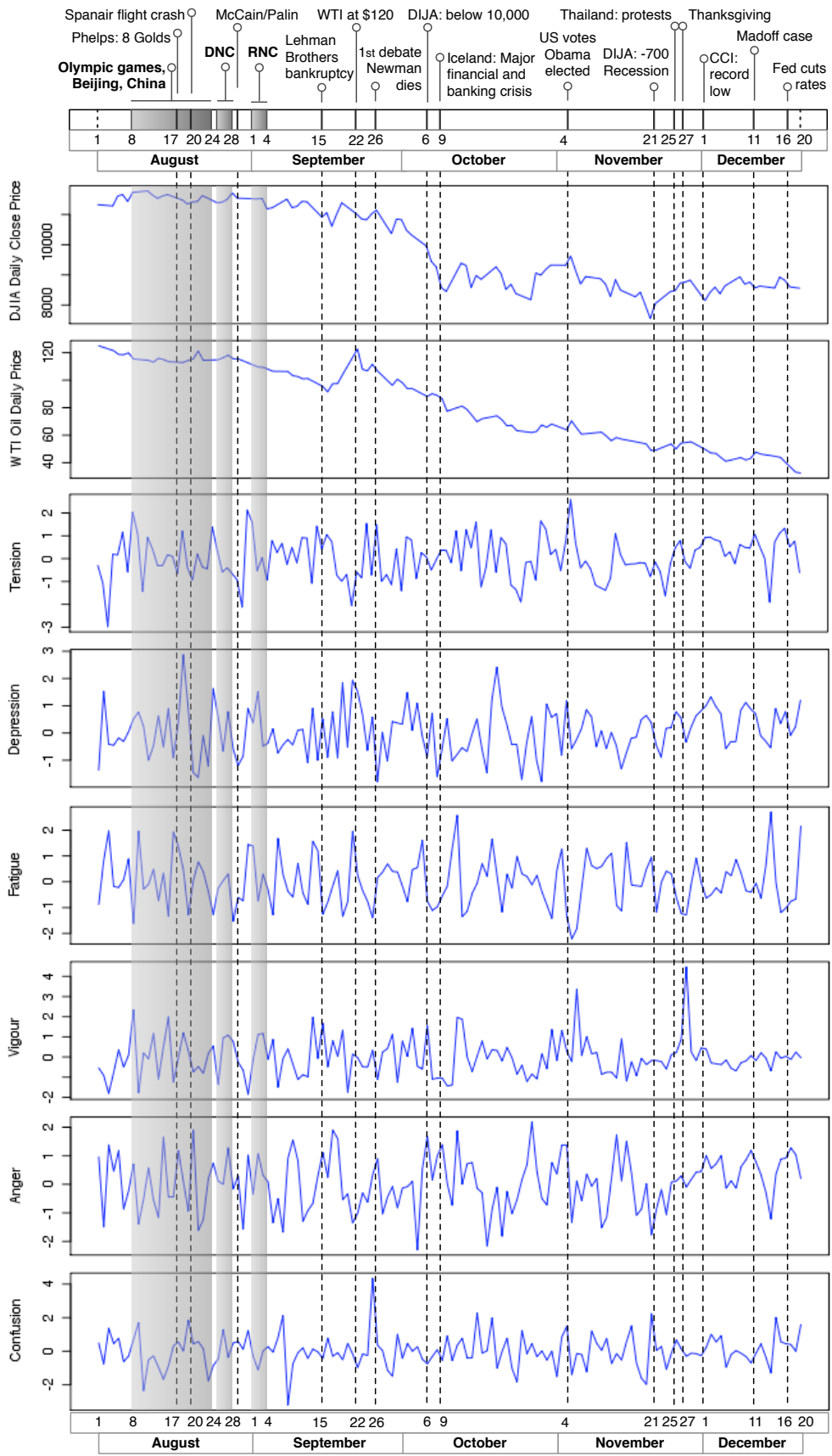

Figure 9: Timeseries for 6 POMS dimensions from August 1 to December 20, 2009. Major events marked in timeline above.